\documentstyle[12pt,psfig,epsf]{article}
\def\href#1#2{#2}
\newcommand{\dd}{{\sf d}}
\begin{document}

\thispagestyle{empty}
 \begin{flushleft}
DESY 98--073
\\
hep-ph/9806465
\\
June 1998
 \end{flushleft}

 \noindent
 \vspace*{0.50cm}
\begin{center}
 \vspace*{2.cm} 
{\huge 
Anomalous Couplings
\vspace*{3mm}
\\
in $W$ Pair Production\footnote{Talk given
at Zeuthen Workshop on Loops and Legs in Gauge Theories, 19--24 April 1998,
Rheinsberg, Germany}}
\\
 \vspace*{2.cm}               

{\large 
Jochen Biebel
}
\vspace*{0.5cm}
 
\begin{normalsize}
{\it
Deutsches Elektronen-Synchrotron DESY Zeuthen,
\\
Platanenallee 6, D-15738 Zeuthen, Germany
}
\end{normalsize}
\end{center}
 
 \vspace*{2.5cm} 

\begin{abstract}
I present a short overview over $W$ pair production and studies of
angular differential cross-sections with and without initial state
radiation applying semi-analytical methods and using the Fortran
program {\tt GENTLE}.
The influence of anomalous couplings to this process is also discussed.
\end{abstract}

\vfill

\section{Introduction}

Since the formulation of the standard model of electroweak
interactions \cite{pSM}, more
and more precision tests have confirmed its validity.
However, for a central part of the theory, the non-abelian structure of
the gauge couplings, we have only poor experimental information.       
With the results of LEP2 \cite{Altarelli:1996ab} and potential
future linear colliders \cite{Accomando:1997wt} the
situation will change.
In processes with $W$ production triple and quartic gauge boson vertices 
appear and may be measured.
Of special interest in $e^+e^-$ annihilation is the process
\begin{equation}
e^+e^-\to W^+W^- ,
\label{Wpairprod}
\end{equation}
with contributions from $\gamma W^+W^-$ and $ZW^+W^-$ vertices.

The current limits for anomalous couplings from combined results of
LEP2 and D0 are \cite{pClare:1998aa}:
\begin{eqnarray}
\alpha_{W\phi}&=&
-0.03^{+0.06}_{-0.06},
\label{limit1}\\
\alpha_W&=&
-0.03^{+0.08}_{-0.08},\\
\alpha_{B\phi}&=&
-0.05^{+0.22}_{-0.20},
\label{limit3}
\end{eqnarray}
where the $\alpha$'s are given by the identities:
\begin{eqnarray}
\alpha_{W\phi}&=&c_Ws_W\delta_Z\vphantom{\frac{c_W}{s_W}},\\
\alpha_W&=&y_\gamma=\frac{s_W}{c_W}y_Z,\\
\alpha_{B\phi}&=&x_\gamma-c_Ws_W\delta_Z=-\frac{c_W}{s_W}
\left(x_Z+s_W^2\delta_Z\right).
\end{eqnarray}
The anomalous parameters $\delta_Z$, $x_\gamma$, $x_Z$, $y_\gamma$,
$y_Z$, and $z_Z$ are defined by the Lagrangian in eq.~(\ref{deflag}).

In section \ref{wpairprod} I give an overview on $W$ pair production
and the studies of the differential cross-sections performed with {\tt
GENTLE} version 2 \cite{Bardin:1996zz}.
I discuss the influence of potential anomalous three gauge boson
couplings to $W$ pair production in section \ref{anocoup}.

\section{$W$ Pair Production}
\label{wpairprod}

The first calculations in the
standard model for the process in (\ref{Wpairprod}) were done
in the narrow width
approximation \cite{pWWonsh
}, e.g. neglecting the finite width of the $W$ bosons.
At this time it was known, that the decay width $\Gamma_W$ of the
$W$ will give rise to large corrections if the $W$ is much heavier
than the proton \cite{Tsai:1965hq}.
As a consequence the finite $W$ width must be considered.
This can be done by convoluting the cross-section with Breit-Wigner
factors \cite{Muta:1986is}:
\begin{equation}
\sigma(s)=\int\limits_0^s\dd
s_1\,\rho(s_1)\int\limits_0^{(\sqrt{s}-\sqrt{s_1})^2}\dd
s_2\,\rho(s_2)\sigma_0(s,s_1,s_2)
\label{muta}
\end{equation}
with
\begin{equation}
\rho(s_i)=\frac{1}{\pi}\frac{\sqrt{s_i}\Gamma_W(s_i)}
                  {(s_i-m_W^2)^2+m_W^2\Gamma_W^2(s_i)}
\times B(f),
\end{equation}
where $s$ is the center-of-mass energy squared and $B(f)$ the branching
fraction for the $W$ decaying in the fermion doublet $f$.
The invariant masses of the decay products of the $W$ bosons are
denoted by $s_1$ and $s_2$.

Since the produced $W$ bosons decay almost immediately, the
production of 4 fermions
\begin{equation}
e^+e^-\to W^+W^-\to 4f
\end{equation}
is observed.
Additional diagrams, so called background diagrams, contribute to
the same final states and should be taken into account, too
\cite{pWWback}.
The number of contributing diagrams depends on the final state
fermions.
Here, I will concentrate on the {\tt CC11} class, since it
includes the semi-leptonic final states. These final states are
important in the measurement of the gauge couplings, because they
offer the most complete kinematical information.
The {\tt CC11} class is defined by having two different
weak doublets and no electrons nor neutrinos as final state fermions.
Depending on the number of produced neutrinos, there are 9, 10, or 11
Feynman diagrams.

For the {\tt CC11} class the $\sigma_0$ of (\ref{muta}) can be written
as a sum over all interferences and combinations of coupling constants.
For the differential cross-section one gets:
\begin{equation}
  \frac{\dd\sigma_0}{\dd\cos\theta}= \frac{\sqrt{\lambda(s, s_{1},s_2)}}{\pi s^2}
  {{ \sum_{k} {\cal C}_k\cdot{\cal G}_k(s;s_1,s_2,\cos\theta)}}
  \label{totsig}
\end{equation}
The coefficient functions ${\cal C}_k$ are rather trivial and contain
the coupling constants of the particles and the $s$-channel
propagators.
The kinematical functions ${\cal G}_k$ are more complicated and
describe the non-trivial dependencies of $s$, $s_1$, and $s_2$ and
other variables like the scattering angle $\cos\theta$.
To express the total cross-section a smaller set of ${\cal C}$ and
${\cal G}$ functions as in eq.~(\ref{totsig}) is needed, since the
parity violating contributions disappear after the integration over
the scattering angle.

As a simple example I give the expressions of the ${\cal C}$ and ${\cal
G}$ functions for the differential cross-section for the square of
the $t$-channel diagram \cite{Bardin:1996uc}:
\begin{equation}
  {\cal C}^t=\frac{\left(G_\mu m_W^2\right)^2}{s_1s_2}
\rho_W(s_1)\rho_W(s_2),
\end{equation}
and
\begin{equation}
  {\cal G}^t=\frac{1}{8}\left[2s(s_1+s_2)+\frac{\lambda}{4}\sin^2\theta
    +\frac{\lambda s_1s_2\sin^2\theta}{t^2_\nu}\right],
\end{equation}
where $\lambda$ is the K\"allen-function
\begin{equation}
\lambda\equiv\lambda(s,s_1,s_2)=s^2+s_1^2+s_2^2-2ss_1-2ss_2-2s_1s_2,
\end{equation}
and $t_\nu$ is the neutrino propagator
\begin{equation}
t_\nu=\frac{1}{2}\left(s-s_1-s_2-\sqrt{\lambda}\cos\theta\right).
\end{equation}
The interferences between signal diagrams and background diagrams are
more complicated and I give only the the kinematical
function for the interference between the $t$-channel diagram and the
$u_1$-diagram as an example:
\begin{eqnarray}
\lefteqn{{\cal G}^{tu_1}(s,s_1,s_2) =}\nonumber\\
&&  \frac{-1}{\lambda}\left\{
  \frac{3}{4}\frac{\cos\theta}{\sqrt{\lambda}}\right.
  s^2s_1s_2^2(5\sin^2\theta-2)\left[\frac{1}{t_\nu}(s+s_1-s_2)
  +2s{\cal L}(s_1;s_2,s)\right]
\nonumber\\&&\mbox{}
  +\lambda\left[\frac{\sin^2\theta}{8t_\nu}[2s_1s_2(s_2-s_1)
  -6s^2s_2(s_1+s_2){\cal L}(s_1;s_2,s)-3ss_2(s+s_2)]\right.
\nonumber\\&&\mbox{}
  \left.+\frac{\sin^2\theta}{16}[(s-s_1)^2-s_2^2]+\frac{ss_1}{2}
  \vphantom{\frac{1}{t_\nu}}\right]
  +\frac{ss_1s_2}{t_\nu}\left[-\frac{3}{4}ss_2{\cal L}(s_1;s_2,s)
  (5s\sin^4\theta
  \right.
\nonumber\\&&\mbox{}
  +4(s_1+s_2))-\frac{1}{8}(3s_2^2-2ss_1+4s_1s_2-7s_1^2+30ss_2+9s^2)
  \sin^2\theta
\nonumber\\&&\mbox{}
  -\frac{1}{2}(3s_2^2-2s_1^2\left.-s_1s_2+2ss_1)\vphantom{\frac{3}{4}}\right]
\nonumber\\&&\mbox{}
  +\frac{3s^2s_2}{4}{\cal L}(s_1;s_2,s)([4s_1s_2+s_1^2+s_2^2-s(s_1+s_2)]
  \sin^2\theta
\nonumber\\&&\mbox{}
  -4[s_1s_2+s_1^2+s_2^2-s(s_1+s_2)])
  +\frac{ss_2\sin^2\theta}{8}(2s_1s_2-5s_1^2+3s_2^2
\nonumber\\&&\mbox{}
  -14ss_1 -3s^2)
  \left.+\frac{s}{2}(5s_1^2s_2-2s_1s_2^2-3s_2^3
  +5ss_1s_2+3s^2s_2)\right\},
\label{tu1}
\end{eqnarray}
with 
\begin{equation}
{\cal
L}(s;s_1,s_2)=\frac{1}{\sqrt{\lambda}}\ln\frac{s-s_1-s_2
+\sqrt{\lambda}}{s-s_1-s_2-\sqrt{\lambda}}.
\end{equation}
The remaining coefficient and kinematical functions are presented in
\cite{Biebel:1998ww}.

\begin{figure}[ht]
  \begin{center}
    \epsfxsize=12.6cm
    \leavevmode
    \epsffile{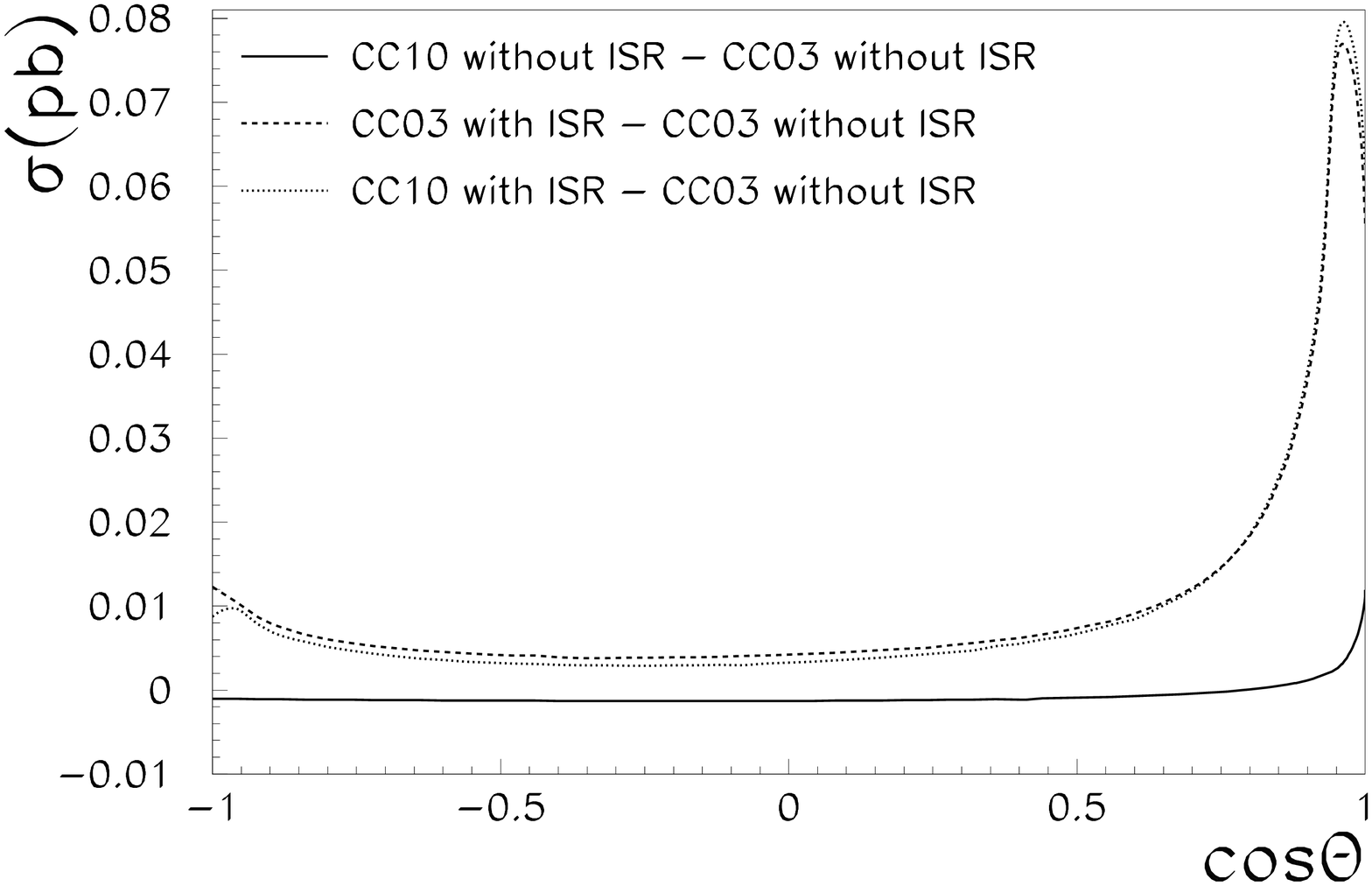}
  \end{center}
  \caption{\label{qed.ps}\it Differences between Born cross-section
  and cross-section with initial state radiation.}
\end{figure}

To make precise predictions for the cross-section, radiative
corrections have to be taken into account \cite{pBeenV}. 
To demonstrate the size of initial state radiation (ISR), I show
in fig.~\ref{qed.ps} the difference between the ISR corrected
cross-section and the Born cross-section in the case of
signal diagrams ({\tt CC03}) and for the complete semi-leptonic
process ({\tt CC10}).    
While the difference peaks in the region $\cos\theta>0.8$ it is
almost constant in the other parts.
Especially in the region of $\cos\theta\to -1$ this leads to important
effects, because the differential cross-section drops here significantly
and the relative corrections amount to 30\% for
$\cos\theta =-1$ \cite{Biebel:1998ww}.

The numerical results of fig.~\ref{qed.ps} were produced with {\tt GENTLE}
and radiative corrections were treated as described in
\cite{Bardin:1996zz}.
A short overview over {\tt GENTLE} is given in appendix \ref{gentle}.

\section{Anomalous Couplings}
\label{anocoup}

The most general form for the $\gamma WW$ and $ZWW$ vertices compatible
with Lorentz invariance was first considered in
\cite{Gaemers:1979hg}, where 9 parameters were introduced for each vertex.
In \cite{Hagiwara:1987vm} it was shown that these parameters were not
independent and the number could be reduced to 7.

The number of parameters can be further reduced by using a 
restricted set of anomalous couplings, which is
invariant under ${\cal CP}$ transformations.
The anomalous couplings are defined by the Lagrangian:
\begin{eqnarray}
  {\cal  L}&=&
  -ie\left[A_\mu\left(W^{-\mu\nu}W^+_\nu-W^{+\mu\nu}W^-_\nu\right) 
  +F_{\mu\nu}W^{+\mu}W^{-\nu}\right]
\nonumber\\&&\mbox{} 
  -ie\left(\cot\Theta_W+\delta_Z\right)\left[Z_\mu\left(W^{-\mu\nu}W^+_\nu
  -W^{+\mu\nu}W^-_\nu\right) 
  +Z_{\mu\nu}W^{+\mu}W^{-\nu}\right]
\nonumber\\&&\mbox{}
-iex_\gamma  F_{\mu\nu}W^{+\mu}W^{-\nu}-iex_ZZ_{\mu\nu}W^{+\mu}W^{-\nu}
\nonumber\\&&\mbox{}
+ie\frac{y_\gamma}{M_W^2}F^{\nu\lambda}W^-_{\lambda\mu}W^{+\mu}_\nu
  +ie\frac{y_Z}{M_W^2}Z^{\nu\lambda}W^-_{\lambda\mu}W^{+\mu}_\nu
\nonumber\\&&\mbox{}
  +\frac{ez_Z}{M_W^2}\partial_\alpha\tilde{Z}_{\rho\sigma}\left(
  \partial^\rho W^{-\sigma}W^{+\alpha}-\partial^\rho
  W^{-\alpha}W^{+\sigma} \right.
\nonumber\\&&\mbox{}\left.
+\partial^\rho W^{+\sigma}W^{-\alpha}-\partial^\rho
W^{+\alpha}W^{-\sigma}\right).
\label{deflag}
\end{eqnarray}
%
In the standard model the anomalous parameters $\delta_Z$, $x_\gamma$,
$x_Z$, $y_\gamma$, $y_Z$, and $z_Z$ are zero. 
The parameter $z_Z$ violates both ${\cal C}$ and ${\cal P}$ symmetry,
but is invariant under the product ${\cal CP}$.
The parameters $x_\gamma$ and $y_\gamma$ contribute to the 
magnetic dipole moment $\mu_W$ and the electromagnetic quadrupole
moment $q_W$ of the $W$ boson \cite{Aronson:1969aa}:
\begin{eqnarray}
\mu_W&=&\frac{e}{2m_W^2}\left(2+x_\gamma+y_\gamma\right),\\
q_W&=&-\frac{e}{m_W^2}\left(1+x_\gamma-y_\gamma\right)
.
\end{eqnarray}

With these additional parameters the cross-section for $W$ pair
production can be written as:
\begin{eqnarray}
   \sigma^{\sf{ano}}&=&\sigma^{\sf SM}\nonumber\\
   &&+~{x_\gamma}\cdot \sigma^{{x_\gamma}}+{x_Z}\cdot\sigma^{{x_Z
}}+\dots
\label{expan}
\\
&&+~{x_\gamma^2}\cdot\sigma^{{x_\gamma} {x_\gamma}}
   +{x_\gamma x_Z}\cdot\sigma^{{x_\gamma
   x_Z}}+{x_Z^2}\cdot\sigma^{{x_Zx_Z}}+\dots\nonumber,
\end{eqnarray}
where the anomalous parameters appear at most bilinearly.

If one considers all anomalous couplings of eq.~(\ref{deflag}) 28
coefficients are needed to calculate $\sigma^{\sf{ano}}$.
Eq.~(\ref{expan}) can also be applied to multi-differential
cross-sections.
In the search for anomalous couplings multi-differential cross-sections
are used, since they contain more kinematical information
\cite{Gounaris:1996rz}.

{\tt GENTLE} can be used to calculate the coefficients in
eq.~(\ref{expan}).
This can be done for the differential cross-section and in the {\tt
CC03} process also for the bin-wise integrated differential
cross-section.
By setting the number of bins to 2, one gets predictions for
forward ($\cos\theta>0$) and backward ($\cos\theta<0$) scattering.
A study of the sensitivity of the forward-backward asymmetry to pairs
of anomalous couplings was performed for the pairs
$(x_\gamma,\delta_z)$ and $(x_\gamma,z_Z)$ in \cite{Biebel:1998ww}
and for $(x_\gamma,x_Z)$ and
$(x_Z,z_Z)$ in \cite{pbieb}.
The forward-backward asymmetry proved to be useful for studies
which include the parity violating parameter $z_Z$. 

\section{Conclusions}

I gave a short report over the present state of {\tt GENTLE} and
the studies of differential cross-sections and anomalous couplings with
it.
It was shown that radiative corrections (initial state radiation)
give sizeable effects to the differential cross-section at a
center-of-mass energy of 500~GeV.
This is especially important in the region of backward scattering,
where the cross-section is small and the corrections are about 30\%.

\bigskip
\noindent
{\large\bf Acknowledgement}\\
I would like to thank the organizers of the conference for their
hospitality and the very pleasant atmosphere they provided.
Further, I am especially grateful to T.~Riemann for the
collaboration on this topic.

\appendix
\section{\tt GENTLE}
\label{gentle}

{\tt GENTLE} version 2 is a Fortran package to calculate
cross-sections for 4 fermion production processes
for charged currents ({\tt CC}) and neutral currents ({\tt NC})
with the semi-analytical method.
Table~\ref{gentab} gives an overview over the different branches of
{\tt GENTLE} and the publications they are based on.
\begin{table}[ht]
\begin{center}
\begin{tabular}{|c|l|l|}
\hline
&QED ISR total cross-section&\cite{Bardin:1993nb}\\
&Background total
cross-section&\cite{Bardin:1996uc}\\
\raisebox{1.5ex}[-1.5ex]{\tt CC}&Anomalous couplings&\cite{Biebel:1998ww}\\
&Differential cross-section
&\cite{Biebel:1998ww}\\
\hline
&QED ISR total cross-section&\cite{Bardin:1996jw}\\
\raisebox{1.5ex}[-1.5ex]{\tt NC}&Background total
cross-section&\cite{Bardin:1995vm}\\
\hline
\end{tabular}
\end{center}
\caption{\label{gentab}\it Overview over the different branches of
{\tt GENTLE}. }
\end{table} 

While in {\tt GENTLE} version 2 the calculation of the differential
cross-section and the effects of anomalous couplings were only
available for the signal diagrams, in the newer version 2.01 these
features were extended to the complete {\tt CC11} class.


\small

\begingroup\begin{thebibliography}{10}

\bibitem{pSM}
S. L. Glashow, {\it Nucl. Phys.} {\bf 22} (1961) 579;\\ S. Weinberg, {\it Phys.
  Rev. Lett.} {\bf 19} (1967) 1264;\\ A. Salam, ``Weak and Electromagnetic
  Interactions'', in {\it Proc. of the Nobel Symposium, 1968, Lerum, Sweden}
  (N. Svartholm, ed.), pp. 367--377, Almqvist and Wiksell, Stockholm, 1968.

\bibitem{Altarelli:1996ab}
G.~Altarelli, T.~{Sj\"ostrand}, and {F. Zwirner (eds.)}, ``Physics at {LEP2}'',
  CERN report CERN 96--01 (1996).

\bibitem{Accomando:1997wt}
{ECFA/DESY LC Physics Working Group} Collaboration, E.~Accomando {\em et~al.},
  {\em Phys. Rept.} {\bf 299} (1998) 1.

\bibitem{pClare:1998aa}
R.~Clare, {``LEP Electroweak Physics Results''}, these proceedings.

\bibitem{Bardin:1996zz}
D.~Bardin, J.~Biebel, D.~Lehner, A.~Leike, A.~Olchevski, and T.~Riemann, {\em
  Comput. Phys. Commun.} {\bf 104} (1997) 161. \\{{\tt GENTLE} is available at
  {\tt http://www.ifh.de/theory/publist.html}}.

\bibitem{pWWonsh}
V.~Flambaum, I.~Khriplovich, and O.~Sushkov, {\em Sov. J. Nucl. Phys.} {\bf 20}
  (1975) 537--540;\\ W.~Alles, C.~Boyer, and A.~J. Buras, {\em Nucl. Phys.}
  {\bf B119} (1977) 125.

\bibitem{Tsai:1965hq}
Y.-S. Tsai and A.~C. Hearn, {\em Phys. Rev.} {\bf 140} (1965) B721--B729.

\bibitem{Muta:1986is}
T.~Muta, R.~Najima, and S.~Wakaizumi, {\em Mod. Phys. Lett.} {\bf A1} (1986)
  203.

\bibitem{pWWback}
F.~A. Berends, R.~Pittau, and R.~Kleiss, {\em Nucl. Phys.} {\bf B424} (1994)
  308--342;\\ D.~Bardin, M.~Bilenky, D.~Lehner, A.~Olchevski, and T.~Riemann,
  {\em Nucl. Phys. (Proc. Suppl.) 37B} (1994) 148--157.

\bibitem{Bardin:1996uc}
D.~Bardin and T.~Riemann, {\em Nucl. Phys.} {\bf B462} (1996) 3--28.

\bibitem{Biebel:1998ww}
J.~Biebel and T.~Riemann, ``{Off-shell W Pair Production with Anomalous
  Couplings: The CC11 Process}'', preprint DESY 98-047 (1998),
  \href{http://xxx.lanl.gov/abs/hep-ph/9805355}{{\tt hep-ph/9805355}}.

\bibitem{pBeenV}
W.~Beenakker {\em et~al.}, ``{$WW$} cross-sections and distributions'', in {\em
  Physics at {LEP2}, {\rm CERN 96--01 (1996)}} (G.~Altarelli, T.~{Sj\"ostrand},
  and {F. Zwirner}, eds.), pp.~79--139, and references therein;\\ W.~Beenakker
  and A.~Denner, these proceedings;\\ A.~Vicini, these proceedings.

\bibitem{Gaemers:1979hg}
K.~J.~F. Gaemers and G.~J. Gounaris, {\em Z. Phys.} {\bf C1} (1979) 259.

\bibitem{Hagiwara:1987vm}
K.~Hagiwara, R.~D. Peccei, D.~Zeppenfeld, and K.~Hikasa, {\em Nucl. Phys.} {\bf
  B282} (1987) 253.

\bibitem{Aronson:1969aa}
H.~Aronson, {\em Phys. Rev.} {\bf 186} (1969) 1434--1441.

\bibitem{Gounaris:1996rz}
G.~Gounaris {\em et~al.}, ``Triple gauge boson couplings'', in {\em Physics at
  {LEP2}, {\rm CERN 96--01 (1996)}} (G.~Altarelli, T.~{Sj\"ostrand}, and {F.
  Zwirner}, eds.), pp.~525--576.

\bibitem{pbieb}
J.~Biebel and T.~Riemann, ``Semianalytic predictions for {$W$} pair production
  at 500 {GeV}''. Contribution to: R. Settles (ed.), Proc. of {\em ECFA/DESY
  Study on Physics and Detectors for the Linear Collider}, DESY 97-123E,
  pp.~139--142, {\tt hep-ph/9709207};\\ J.~Biebel, ``{Four fermion production
  with anomalous couplings at LEP-2 and NLC}'', talk at {\it XIIth
  International Workshop on High Energy Physics and Quantum Field Theory, 4--10
  Sep~1997, Samara, Russia}, DESY 97-219 (1997), {\tt hep-ph/9711439}.

\bibitem{Bardin:1993nb}
D.~Bardin, A.~Olshevsky, M.~Bilenky, and T.~Riemann, {\em Phys. Lett.} {\bf
  B308} (1993) 403--410. {\rm E: ibid., B357 (1995) 725}, hep-ph/9507277.

\bibitem{Bardin:1996jw}
D.~Bardin, D.~Lehner, and T.~Riemann, {\em Nucl. Phys.} {\bf B477} (1996)
  27--58.

\bibitem{Bardin:1995vm}
D.~Bardin, A.~Leike, and T.~Riemann, {\em Phys. Lett.} {\bf B344} (1995)
  383--390.

\end{thebibliography}\endgroup
\end{document}